\documentclass[12pt, draftclsnofoot, onecolumn]{IEEEtran}
\IEEEoverridecommandlockouts
\usepackage{cite}
\usepackage{graphicx}
\usepackage{amsmath}
\usepackage{amssymb}
\usepackage{algorithmicx}
\usepackage{algorithm}
\usepackage{color}
\usepackage{bm}
\usepackage{setspace}

\allowdisplaybreaks

\begin{document}

\title{Integrated Sensing and Communication with Multi-Domain Cooperation}

\author{
Jie Yang, Xi Yang, Chao-Kai Wen, and Shi Jin
\thanks{\setlength{\baselineskip}{9pt} Jie~Yang, Xi~Yang, and Shi~Jin are with the National Mobile Communications Research Laboratory, Southeast University, Nanjing, China (e-mail: \{yangjie;yangxi;jinshi\}@seu.edu.cn). Chao-Kai~Wen is with the Institute of Communications Engineering, National Sun Yat-sen University, Kaohsiung, 804, Taiwan (e-mail: chaokai.wen@mail.nsysu.edu.tw).}
}

\maketitle	

\begin{abstract}
	With the continuous increase of the spectrum and antennas, endogenous sensing is now possible in the fifth generation and future wireless communication systems.
	However, sensing is a highly complex task for a heterogeneous communication network with massive connections.
	Seeking multi-domain cooperation is necessary.
	In this article, we present an integrated sensing and communication (ISAC) system that performs active, passive, and interactive sensing in different stages of communication through hardware and software. 
	We also propose different methods about how multi-user and multi-frequency band cooperate to further enhance the ISAC system's performance.
	Finally, we elaborate on the advantages of multi-domain cooperation from the physical layer to the network layer for the ISAC system.

\end{abstract}

\vspace{0cm}
\section{Introduction}
The future wireless communication system is expected to provide ubiquitous connectivity with ultra-high throughput, ultra-low latency, and ultra-high reliability. It will also go far beyond communication to realize the ability to sense, control, and even optimize wireless environments \cite{6G0}. Because, with the rapid advent of the intelligent age, more and more new applications emerge out, such as extended reality, holographic communication, autonomous driving, smart medical, intelligent industry, \emph{etc.}, which require mass data transmission, centimeter-level localization, and highly fine-grained environmental information. To bring the applications above into reality and fulfill the vision that everything will be sensed, connected, and intelligent, integrated sensing and communication (ISAC) is anticipated to play a pivotal role.

The continuous evolution of wireless communication is conspicuously witnessed. On the one hand, millimeter wave (mmWave) and even Tera-Hertz (THz) frequencies are further exploited \cite{locsum}, which coincides with the spectrum of mmWave radar and high-resolution THz imaging radar. Large bandwidth brings high time resolution. On the other hand, the antenna deployment is further scaled up, which enables unprecedented angular resolution \cite{J1}. Coupled with ultra-dense network deployment, the future wireless communication system is enabled a paradigm shift in sensing capabilities. Moreover, reconfigurable intelligent surfaces \cite{wu,tang}, artificial intelligence \cite{DoM,J2}, and integrated terrestrial-aerial-satellite networks \cite{UAV} are regarded as promising techniques to boost both sensing and communication performance. 
It is feasible to integrate the advanced technologies by sharing the infrastructure and time-frequency-space resources of wireless communications.
Integrating sensing and communication functions into one hardware system contributes to reducing cost, power consumption and deployment complexity.
The cooperation and mutual assistance of the two functions can be conveniently achieved by utilizing the high-throughput and low-latency information sharing ability of the wireless communication systems \cite{loc}. Therefore, ISAC is the future trend of technology development, which breaks the traditional pattern of the isolated design of communication and sensing.
ISAC can build a bridge between the physical and digital worlds and lay a solid foundation for the intelligence of everything. 

Although considerable advantages of ISAC have been foreseen, deep integration of sensing and communication still has a long way to go. Start from the physical layer, frame structure should be designed, and the time-frequency-space resources should be allocated to make communication and sensing share the same physical layer processing. Moreover, it is difficult for a single network, not to mention a single user, to complete such a complex task of realizing high-quality communication and high-accuracy sensing \cite{co0,co1}. Multi-domain cooperation can provide different observations of common targets, resulting in more accurate sensing decisions. Local sensing results can be combined to synthesize more comprehensive global results through multi-domain cooperation. Multi-domain cooperation can also provide an effective way to overcome the weaknesses of a single network or user. For example, although the mmWave massive multi-input multi-output network has high sensing resolution, its coverage is limited, sub-6 GHz can ensure reliable links to assist mmWave, wireless local area networks (WLANs) and sensor networks can also provide side information to mmWave networks \cite{co2,co3}.
Therefore, it is of paramount significance to develop flexible and efficient cooperation among multi-domain. Correspondingly, a protocol interface and information fusion scheme should be developed to enable information sharing across users, access points, and even heterogeneous networks. 

In this article, we propose a series of methods to enhance the ISAC in the perspective of multi-domain cooperation, including various sensing types cooperation with the corresponding signal processing and physical layer design, crowdsourcing-based multi-user cooperation with the corresponding information fusion and sharing schemes, and different frequencies cooperation in heterogeneous networks with the corresponding network topology design and protocol interface development. The proposed methods also concerning how the sensing result can be used to speed up the establishment of communication links and reduce the overhead of beam training and channel sounding.

\section{Fundamentals of the ISAC System}

In this section, we introduce the fundamental concepts of the ISAC system. The sensing function of the ISAC system means obtaining the direction, distance, velocity, position, and even image of the target and map of the propagation environment. Simultaneous localization and mapping (SLAM) is one of the promising techniques to achieve the sensing function in the ISAC system. SLAM mainly exploits the unchanging nature of the position and state of radio features in the propagation environment. Through accumulating measurements in consecutive time slots, each radio feature can be mapped to a geometric place (i.e., sensing); then, the mapped results assist the localization of the moving user, thereby extending a single-shot localization solution to the multiple-shot SLAM \cite{slam}.

\begin{figure}
	\centering
	\includegraphics[scale=0.55]{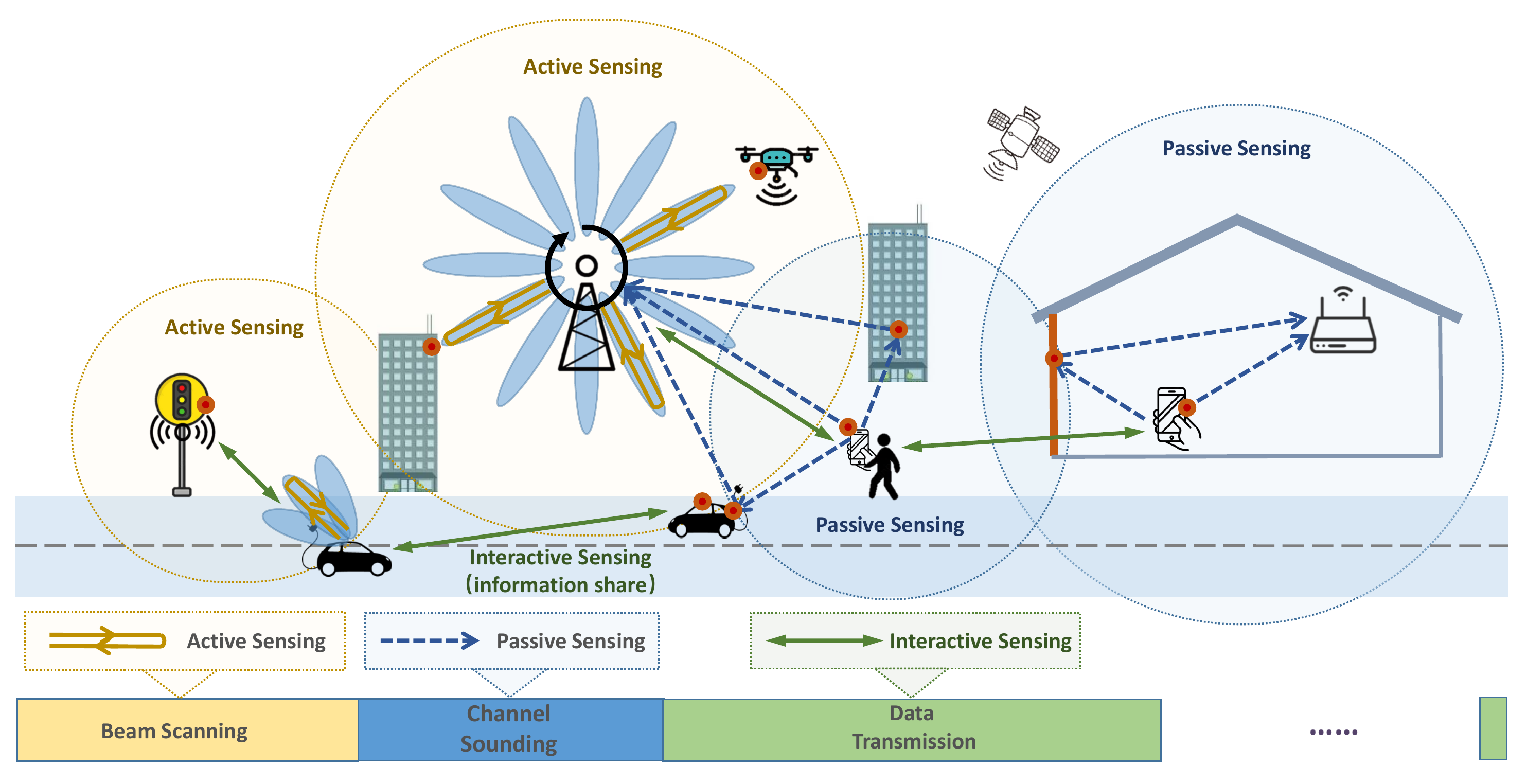}
	\caption{Illustration of ISAC systems in the heterogeneous network. Active, passive, and interactive sensing correspond to beam scanning, channel sounding, and data transmission in communication processes, respectively.}
	\label{system} 
\end{figure}
In general, we can classify the sensing in the ISAC system into three types, namely, active, passive, and interactive types, as shown in Fig. \ref{system}. For active type, the system sends radio frequency (RF) signals and then captures targets from the echo signals.
The system serves as a conventional radar with sensing ability in the active type.
For passive-type, the system does not send sensing signals while capturing targets through the received RF signals sent or reflected from the targets. 
The system functions as SLAM in the passive type.
For interactive-type, position and environment information is transmitted among transceivers during the data transmission stage.
The system acts as a traditional communication system to share information in the interactive type.
Existing research focuses on a single type. In fact, the three sensing types can be simultaneously achieved. 
Specifically, as shown in Fig. \ref{system}, active sensing, passive sensing, and interactive sensing may correspond to beam scanning, channel sounding, and data transmission in the communication processes, respectively. 
One of the fundamental problems of ISAC is how to allocate the resource between sensing and communication. Therefore, theoretical guidance of the trade-off between the performance and the resource should be established.


In the era of the internet of everything, the ISAC system should provide services for a large number of users, including vehicles, mobile phones, internet of things devices, \textit{etc.} These users can also act as sensing result providers, thus forming a framework for crowdsourcing. Through multi-user cooperation, the ISAC system will be cost-effective. In addition, the ISAC system is expected to involve satellite, unmanned aerial vehicle (UAV), radar, cellular, WLAN, and sensor networks, which together form a vast heterogeneous network (Fig. \ref{system}). Each kind of network has its advantages and disadvantages. For example, satellite-based global positioning system (GPS) can provide position information outdoors, but it is ineffective indoors; WLAN can be economically deployed indoors, whereas its sensing resolution is limited. Therefore, cooperation among heterogeneous networks is imperative.

The ultimate goal of the ISAC system is in two aspects. On the one hand, the cellular system obtains new functions, which can track devices, localize targets, identify objects, and map radio environments. On the other hand, communication performance is enhanced by cellular sensing.
In particular, we aim to improve the sensing performance with limited resources, and then accurate sensing results help reduce the communication overhead in beam training and channel sounding. For example, when the transceiver positions and the multi-path propagation environment are known, the effective beam directions can be predicted, which speeds up establishing the mmWave communication links. Therefore, we put forward new ideas for the physical layer and network layer design in ISAC systems, the benefits and challenges are explained in detail in the following section.

\section{ISAC System with Multi-domain Cooperation}\label{cooperation}

We present a novel ISAC system with multi-domain cooperation, such as cooperation through active and passive sensing, multi-user, and multi-frequency band networks.
With the proposed ISAC system, diverse sensing types, numerous terminals, and heterogeneous networks can improve each other.

\subsection{Multi-Sensing Type Cooperation}\label{active_passive}
\begin{figure}
	\centering
	\includegraphics[scale = 0.45]{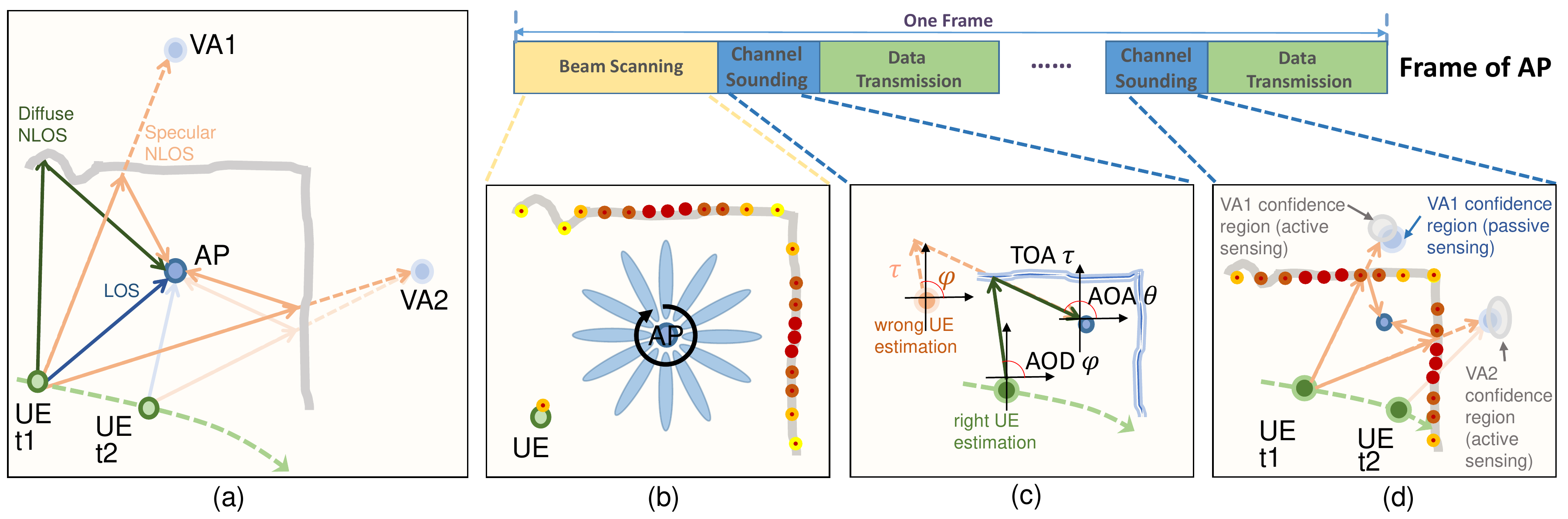}
	\caption{The active and passive combined ISAC system with the corresponding frame structure.
		(a)	illustrates the radio feature map, where the gray lines, blue dots, and green dots represent the scatterers, AP and the corresponding VAs, and UE positions, respectively.
		(b) presents the process of active sensing, where red and yellow dots denote the mapped scatterers.
		(c) shows an example of active and passive sensing cooperation in the diffuse reflection scenario, where the blue line is the scatterer shape obtained from interpolation, the green dot denotes the unique solution of UE position, and the red dot is a wrong UE position estimate.
		(d) gives an example of active and passive sensing cooperation with specular reflection. The gray ellipse and blue dot denote the VA confidence region of active and passive sensing, respectively.}
	\label{A1} 
\end{figure}
Fig.~\ref{A1} shows the frame structure of the proposed ISAC system. In the beam scanning stage, the access point (AP) sweeps a bank of discrete beam directions with RF signal and the user equipment (UE) detects the amplitudes of each beam to determine the proper transmission and receive beams.
Meanwhile, the AP can listen to the echo signals to accomplish active sensing.
For example, the time-of-arrival (TOA) of each beam direction can be estimated from the echo signals. Thus, the surrounding scatterers around the AP can be built. In the channel sounding stage, the UE may transmit the pilot signals from several beam directions. The AP can extract the propagation parameters such as angle-of-arrival (AOA), angle-of-departure (AOD), and TOA of propagation paths from the received signal.
With the estimated propagation parameters, the AP can perform passive sensing by localizing the UE and mapping the scatterers between UE and AP. Note that the UE can also perform beam scanning to sense the surrounding radio environment actively. With the cooperation between active sensing and passive sensing, a radio feature map with UE position can be charted. In particular, the legacy scatterers on the map will be refined, and the new scatterers will be added to the map over time. The radio feature map not only enables various UE position-based applications but also improves communication. For example, the feasible UE position region of the next time slot can be predicted through that of the current time slot with the velocity and acceleration. Therefore, the overhead of beam scanning can be reduced by limiting the scanning range to a few possible beam directions.

An instance of active sensing and passive sensing is depicted in Fig.~\ref{A1}(a) with line-of-sight (LOS) path and diffuse and specular non-line-of-sight (NLOS) reflected paths, where the active sensing is performed at AP. In Fig.~\ref{A1}(a), the virtual anchor (VA) represents the mirror of AP with respect to the scatterer \cite{slam}. Because the impact of high-order reflections can be neglected due to the severe attenuation, only the LOS and the first-order NLOS reflected paths are considered. For active sensing, the scatterers that surround the AP are mapped through multiple scanning beams, which in total cover 360 degrees (for 2-dimensional case). In Fig.~\ref{A1}(b), the dots represent the mapping results, and the dot with darker color corresponds to the echo signal with a larger amplitude which implies the better scatterer estimate. Clearly, active sensing shows the panorama of the radio environment around AP. In contrast, passive sensing only obtains the radio environment of a few propagation paths, as shown in Fig.~\ref{A1}(a). Passive sensing is capable of providing gridless AOAs and AODs of propagation paths, which implies better UE localization performance through advanced spectral estimation algorithms \cite{ce3}, whereas the UE localization performance of active sensing is limited because the beam directions are restricted to a set of grid points. The SLAM can be effectively realized by the cooperation between active and passive sensing. 
Next, we present two cooperation methods and the corresponding advantages for scenarios with diffuse and specular reflections.

Fig.~\ref{A1}(c) illustrates a general cooperation method regardless of diffuse and specular reflections, where the active sensing provides the scatterer position to assist the UE localization in passive sensing. During the beam scanning stage, a series of discrete reflection points are provided by active sensing, and the continuous scatterer shape can be obtained through interpolation. Furthermore, the association of the reflection points of the current time slot and that of the previous time slots on the radio feature map yields more accurate scatterer shape. Given the scatterers and the estimates of propagation parameters (AOD~$\varphi$, AOA~$\theta$, and TOA~$\tau$) of diffuse NLOS path obtained from the channel sounding stage, the unique solution of UE position can be estimated. However, the unique solution does not exist if the scatterer position is unknown, where a wrong UE estimation is obtained in Fig.~\ref{A1}(c). According to the UE and AP positions, the reflection point at the scatterer of each propagation path can be obtained and saved in the radio feature map.

Another cooperation method can be used to enhance the performance of SLAM by using specular reflections. 
As shown in Fig.~\ref{A1}(d), the UE, the VA, and the reflection point are collinear according to the law of reflection.
The AP and VA remain constant even when the UE moves.
Given VA, the reflection point at scatterer can be obtained accordingly using the law of reflection. 
Therefore, the environment sensing is equivalent to find the VA.
The scatterer shape is interpolate by the discrete reflection points obtained by active sensing, and the corresponding VA can be  obtained according to the law of reflection.
Specifically, those reflection points that belong to any legacy VA are used to refine the estimate of this VA through data fusion.
New VAs are obtained from the remaining reflection points.
Next, in the channel sounding stage, the multiple paths with extracted propagation parameters are associated with the legacy VAs. The SLAM performs the following steps sequentially: 1) UE position: the LOS path and those reflected paths with associated VA are fused to estimate the UE position. 2) Legacy VA refinement: after obtaining the UE position, the legacy VA of each reflected path is refined. 3) New VA: the new VAs are estimated using the remaining (unused) reflected paths. 

Many topics are wide open for the cooperation between active sensing and passive sensing. First, the diffuse and specular reflections should be distinguished in order to use the appropriate cooperation method. Second, the design of high-performance and low-complexity algorithms, such as the association algorithm between the current scatterers/paths and the legacy scatterers/VAs, the fusion algorithm of multiple paths for UE localization, and the fusion algorithm of current and legacy scatterers/VAs for mapping, are fundamental and critical for SLAM.

\subsection{Multi-User Cooperation}\label{crowdsourcing}
The connectivity of humans and machine continues to increase in the age of the internet of everything.
Thus, the terminals are expected to share the information for cooperation. However, the potential benefits provided by multiple terminals are not fully utilized in the conventional system. As discussed in Section~\ref{active_passive}, each UE can sense the local radio environment of the small surrounding area in the communication process and charts the local radio feature map in the ISAC system by active and passive sensing. 
Because each UE can provide position and trajectory information, the global radio feature map of a large area can be quickly established and refined using the data collected by multiple UEs. For the UEs that have overlapped sensing areas, the cooperation helps each other to refine the shared part of the local radio feature map. Note that the sensing performance is closely related to the relative positions among the UE, the scatterer, and the AP. If some UEs detect a scatterer with high confidence level, other UEs that observe this scatterer can directly inherit the reliable scatterer estimate.

\begin{figure} 
	\centering
	\includegraphics[scale = 0.42]{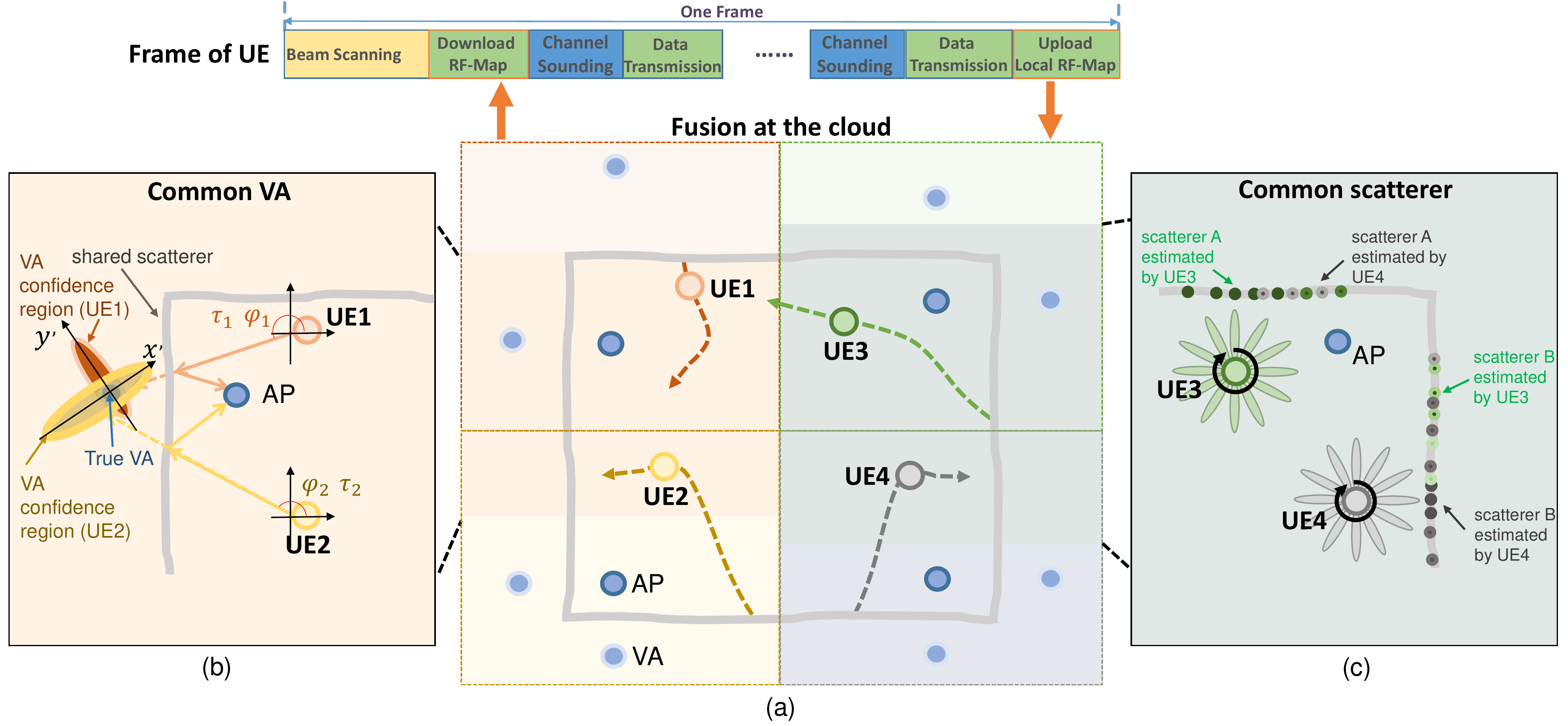}
	\caption{The crowdsourcing-based ISAC system and the corresponding frame structure. 
		(a) shows the global radio feature map (RF-Map), which is fused by four local RF-Map provided by users with different positions and trajectories.
		(b) presents an example of cooperation in the specular reflection scenario, where different users share a common VA.
		(c) expresses an example of cooperation in the diffuse reflection scenario, where different users share common scatterers.}
	\label{B1} 
\end{figure}
As illustrated in Fig.~\ref{B1}(a), the local radio feature maps obtained from four UEs catch a subset of radio features (VAs or scatterers) in the region of interest (ROI). By uploading the local radio feature maps to the cloud database, a global radio feature map can be built by fusing the local maps. Because multiple UEs can sense a common ROI, the radio feature that only exists in a short time is more likely to be caught and recorded from the cooperation, and the labor-intensive data collection is circumvented in the multiuser communication. Although the radio features of each local map may have a low confidence level, the fusion of the crowdsourcing data of a shared radio feature (recorded by different maps) can obtain a refined radio feature with high reliability in the global map. Then, the refined radio features of the global map can be downloaded by the UEs. Each UE continuously refines the legacy radio features, adds the new radio features, and reports the local map to the cloud database, while the cloud refines and distributes the shared features to the corresponding UEs. In addition, each UE can actively detect a radio feature to check whether it still exists or not. If a radio feature vanishes, UE can report to the cloud to delete this radio feature in the global map. For a newly accessed UE, its local map can be initialized by downloading the selected part of the global map to reduce sensing overhead. We present the cooperation methods for a common feature shared by different UEs in the following discussions.

First, we consider the case with common VAs.
According to the discussions in Section~III-A, the VA describes the characteristics of scatterer for specular reflection. In Fig.~\ref{B1}(b), UE~1 and UE~2 share a common VA (scatterer). UE~1 and UE~2 provide different observations to the shared VA from different positions. Assuming that the AOD estimates ${\hat \varphi}_1$ and ${\hat \varphi}_2$ and the TOA estimates ${\hat \tau}_1$ and ${\hat \tau}_2$ of paths~1 and 2 are extracted from the received signals. According to the studies about the multipath channel sounding \cite{J2}, the extracted parameter is generally a Gaussian-corrupted observation of true value such as ${\hat \varphi}_1 = \varphi_1 + w$, where $w$ is Gaussian noise with zero mean and certain variance. Therefore, ${\hat \varphi}_1$ and ${\hat \tau}_1$ form a 2-dimensional Gaussian-corrupted observation of VA position. According to the properties of Gaussian distribution, the confidence region of VA is represented by an ellipse. In practice, the resolutions of AOD and TOA are usually different. Fig.~\ref{B1}(b) describes the case that the TOA measurements are more accurate than the AOD measurements, which means the accuracy in $x'$-axis is superior to that of $y'$-axis for UE~1. Moreover, ${\hat \varphi}_2$ and ${\hat \tau}_2$ form another 2-dimensional Gaussian-corrupted observation from the perspective of UE~2, where the accuracy of $y'$-axis is superior to that of $x'$-axis. Hence, the data fusion of the two confidence regions of the VA takes the advantages of both sides and yields a much more accurate VA estimate.

Next, we consider the case with common scatterer.
Fig.~\ref{B1}(c) shows an instance of the propagation environment, where UE~3 and UE~4 perform beam scanning to sense the local radio environment. As shown in Fig.~\ref{B1}(c), UE~3 and UE~4 share two common scatterers~A and B. Each UE observes a series of reflection points, and the cooperation between UE~3 and UE~4 can improve the sensing performance of both UE. Note that the distance between UE~3 and scatterer~A is much smaller than that between UE~4 and scatterer~A. This implies that the sensing performance of scatterer~A provided by UE~3 is much better than that provided by UE~4 because a smaller distance means larger echo signal amplitude. Therefore, the scatterer~A on the local radio feature map of UE~4 can be improved using the local map of UE~3. Similarly, the information of scatterer~B provided by UE~4 can help UE~3 to refine its local estimate about scatterer~B. Although the local information is incomplete and biased, the cooperation helps each UE obtain comprehensive global information. The abovementioned examples show that the cooperation refines the shared information of each UE, which significantly speeds up the mapping of the global radio feature map.

One of the main challenges for the multi-user cooperation is to measure the confidence level (reliability) of detected radio features. According to the confidence level, unreliable radio features can be filtered out to reduce data processing overhead. In addition, a good pruning algorithm should be designed to reduce the amount of radio map data transmitted between UE and cloud to save the overhead of wireless fronthaul and backhaul.

\subsection{Multi-Frequency Band Cooperation}\label{heterogeneous}

In Sections~III-A and III-B, we present the fundamental localization and mapping methods and the corresponding cooperation strategies in the physical layer. As discussed above, the different perspectives provide distinct but incomplete and biased side information, and the cooperation of side information shows the possibility of revealing the panorama of global information. Another strategy is to fuse the heterogeneous networks, such as sub-6 GHz and mmWave cellular systems, WLANs, sensor networks, UAVs, satellites, radar systems, {\em etc.}, in the network layer to realize the multi-frequency band cooperation. The performance of each network can be improved by taking advantage of multi-domain information. In the following discussions, we present the cooperation method for sub-6 GHz and mmWave systems.

\begin{figure}
	\centering
	\includegraphics[scale = 0.395]{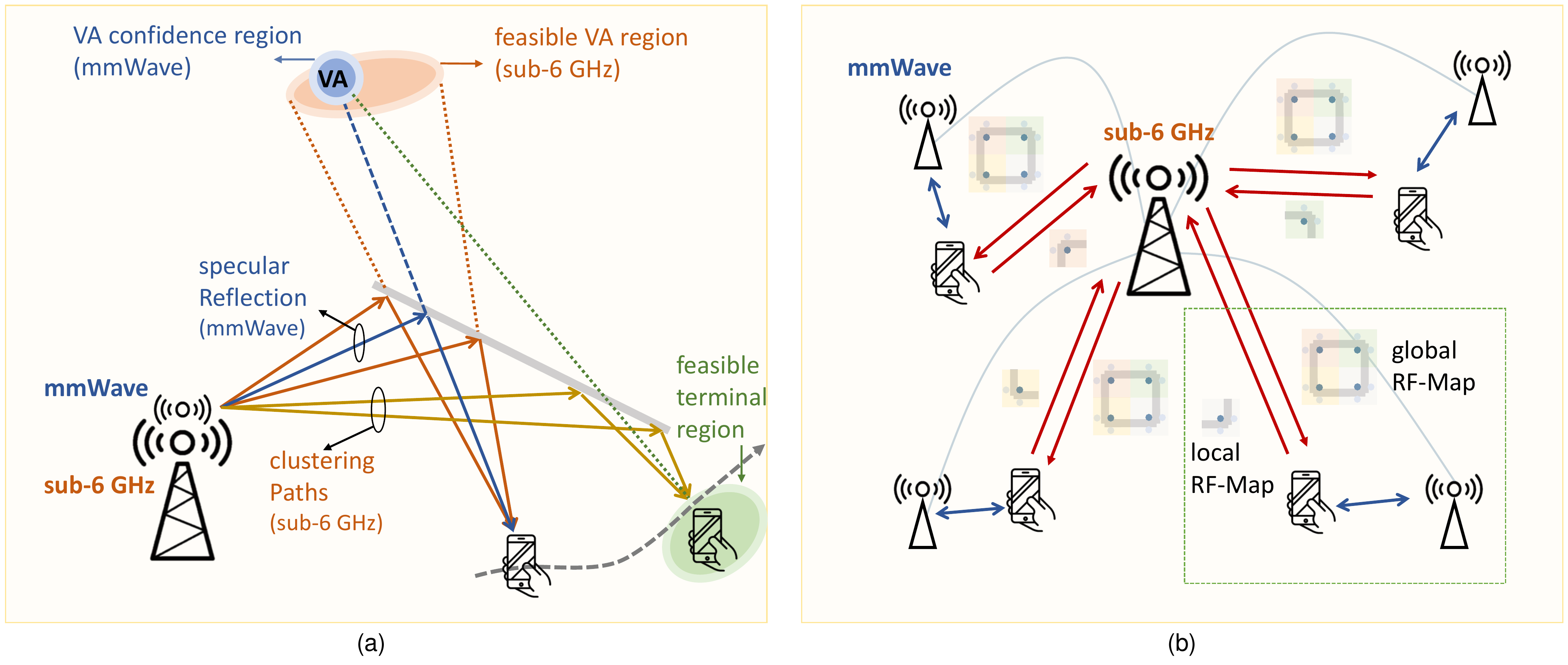}
	\caption{The cooperation of sub-6 GHz and mmWave in the ISAC systems. (a) presents an example of the collocated deployment. (b) provides an instance of the distributed deployment. }
	\label{C1} 
\end{figure}

The mmWave system offers a high data rate using large bandwidth in high frequency.
However, the severe attenuation and shadowing imply poor quality of service. Compared with the mmWave communication, the sub-6 GHz system provides broad coverage and reliable link quality. 
The future cellular system is expected to contain both sub-6 GHz and mmWave APs, where the sub-6 GHz APs ensure seamless communication, and the mmWave APs provide high-quality access in hot spot areas. The latest release of 5G New Radio (NR) supports dual-connectivity, where the mobile terminal is able to connect multiple nodes such as eNB for Long Term Evolution (LTE) and gNB for 5G NR. There are two types of deployment strategies for sub-6 GHz and mmWave APs: 1)~collocated deployment and 2)~distributed deployment. For collocated deployment, the sub-6 GHz and mmWave APs are located in the same position, where the former mainly covers the cell-edge terminals, and the latter supports high data rate transmission for cell-center terminals. The sub-6 GHz and mmWave APs are located in different positions in a distributed deployment. The mmWave AP serves the local terminals in the hot spot area as micro AP and the sub-6 GHz AP provides reliable access to the terminal as macro AP. Considering the propagation characteristics, different cooperation methods should be used for different deployments.

For collocated deployment, the sub-6 GHz and mmWave APs share some common scatterers. However, the propagation phenomenon via shared scatterer is different in sub-6 GHz and mmWave communications. In practice, the sub-6 GHz usually observes a bundle of clustering paths, whereas the multipath in mmWave is much sparser, and the specular reflection can be observed. As shown in Fig.~\ref{C1}(a), for the common scatterer, mmWave transceivers only observe a single specular reflection path that associates to a VA, whereas sub-6 GHz transceivers obtain a batch of clustering paths. The means and the spreads of the AOA, AOD, and TOA of clustering paths with respect to the shared scatterer can be extracted from the received signal at sub-6 GHz transceivers. In Fig.~\ref{C1}(a), a feasible VA region is obtained using the means and the spreads of extracted parameters. According to the feasible VA region, the overhead of beam scanning in mmWave communication can be considerably reduced. This approach is especially suitable for a newly accessed terminal, where the local radio feature map is not yet charted, and the reliable radio environment information is unavailable. After the initial access, the accurate VA estimate is obtained through mmWave communication and recorded on the local radio feature map. Given the VA, if the link between the terminal and mmWave AP undergoes severe attenuation or shadowing, the terminal can switch to the sub-6 GHz communication and use the extracted parameters to obtain a feasible terminal region. Once the link quality between the terminal and mmWave AP is available, the terminal can switch to the mmWave communication and use the feasible terminal region to reduce the overhead of beam scanning.

In the distributed deployment, each mmWave AP serves the terminals in the local hot spot area and establishes the local radio feature map. The sub-6 GHz AP has broad coverage and reliable link quality, taking charge of the up-loading/down-loading of the radio feature map and the network control. As shown in Fig.~\ref{C1}(b), the local radio feature maps of mmWave transceivers can be uploaded to the sub-6 GHz AP, the accurate position and trajectory of the terminal obtained by mmWave transceivers help the localization of sub-6 GHz AP and VAs. 
The sub-6 GHz AP can also generate the global radio feature map by fusing the local maps and distribute the selected radio features to each terminal using the wireless fronthaul. In particular, when a terminal enters the service area of the heterogeneous network, sub-6 GHz AP can quickly establish the communication link and distribute the local radio feature map to the terminal.  
The terminal can quickly access the local mmWave AP by matching its rough position obtained from GPS with the downloaded radio feature map.
In addition, only the selected radio features of the local map with respect to the terminal position are transmitted to avoid the overhead of wireless fronthaul. If the terminal is successfully access to the mmWave AP, the subsequent map downloading can be completed through mmWave communication.

In the above discussions, we only present two possible cooperation methods for sub-6 GHz and mmWave systems. It is critical to analyze the specific correlations between various communication and sensing systems and design a general data fusion algorithm framework to fuse the correlated information from multi-domain effectively. In addition, the radio features which may contain the UE position information are inevitably uploading and downloading through the wireless fronthaul and backhaul. Hence, how to protect privacy is a crucial problem for the ISAC system.

\section{Conclusions and Future Directions}

In this article, we have presented a novel ISAC system that performs active, passive, and interactive sensing in different stages of communication through hardware and software.
We also elaborated on how multi-domain cooperation can be achieved to enhance the performance of the ISAC system through active and passive sensing, multi-user, and multi-frequency band networks.

The ISAC system with RIS and AI are potential research directions.
On the one hand, RIS can manipulate electromagnetic waves in a cost-effective and energy-efficient manner. 
Therefore, RIS promises to extend the wireless communication range, facilitate NLOS communications, and provide low-cost cooperative localization and sensing opportunities.
The ISAC system with RIS assistance can open another dimension in environment sensing.
On the other hand, AI has good learning ability, which can extract useful features from massive data and alleviate modeling issues of ISAC systems in complex environments.
Hybrid data and model-based AI methods can reduce the training overhead and enhance the robustness of the ISAC systems.

\end{document}